\documentclass[aps,letterpaper,twocolumn,prl,showpacs,nofootinbib]{revtex4-1}
\usepackage{amssymb}
\usepackage{amsmath,bm}

\usepackage[pdftex]{graphicx,color}
\usepackage[english]{babel}
\usepackage{hyperref}
\usepackage{enumitem}
\usepackage{epstopdf}
\usepackage{tikz}

\begin{document}

\title{Spontaneous symmetry breaking of dissipative optical solitons in a two-component Kerr resonator}

\author{Gang Xu$^{1,2}$}
\author{Alexander Nielsen$^{1,2}$}
\author{Bruno Garbin$^{1,2,3}$}
\author{Lewis Hill$^{4,5}$}
\author{Gian-Luca Oppo$^{4}$}
\author{Julien Fatome$^{6}$}
\author{Stuart G. Murdoch$^{1,2}$}
\author{St\'ephane Coen$^{1,2}$}
\author{Miro Erkintalo$^{1,2}$}
\email{m.erkintalo@auckland.ac.nz}
\affiliation{$^1$Department of Physics, University of Auckland, Auckland 1010, New Zealand}
\affiliation{$^2$The Dodd-Walls Centre for Photonic and Quantum Technologies, New Zealand}
\affiliation{$^3$Centre de Nanosciences et de Nanotechnologies (C2N), CNRS, Université Paris-Saclay, F-91120 Palaiseau, France}
\affiliation{$^4$SUPA and Department of Physics, University of Strathclyde, Glasgow G4 0NG, Scotland, EU}
\affiliation{$^5$National Physical Laboratory, Hampton Road, Teddington, TW11 0LW, UK}
\affiliation{$^6$ICB, UMR 6303 CNRS, Université Bourgogne-Franche-Comté, 9 Av. Alain Savary, BP 47870, F-21078 Dijon, France}

\begin{abstract}
Dissipative solitons are self-localised structures that can persist indefinitely in ``open'' systems characterised by continual exchange of energy and/or matter with the environment. They play a key role in photonics, underpinning technologies from mode-locked lasers to microresonator optical frequency combs. Here we report on the first experimental observations of spontaneous symmetry breaking of dissipative optical solitons. Our experiments are performed in a passive, coherently driven nonlinear optical ring resonator, where dissipative solitons arise in the form of persisting pulses of light known as Kerr cavity solitons. We engineer balance between two orthogonal polarization modes of the resonator, and show that despite perfectly symmetric operating conditions, the solitons supported by the system can spontaneously break their symmetry, giving rise to two distinct but co-existing vectorial solitons with mirror-like, asymmetric polarization states.  We also show that judiciously applied perturbations allow for deterministic switching between the two symmetry-broken dissipative soliton states, thus enabling all-optical manipulation of topological bit sequences. Our experimental observations are in excellent agreement with numerical simulations and theoretical analyses. Besides delivering fundamental insights at the intersection of multi-mode nonlinear optical resonators, dissipative structures, and spontaneous symmetry breaking, our work provides new avenues for the storage, coding, and manipulation of light.
\end{abstract}

\maketitle

\section{Introduction}

\noindent Temporal cavity solitons (CSs) are the dissipative optical solitons~\cite{akhmediev_dissipative_2008,ackemann_chapter_2009, grelu_dissipative_2012} of coherently-driven nonlinear resonators~\cite{wabnitz_suppression_1993}. First observed in a macroscopic optical fibre ring resonator~\cite{leo_temporal_2010}, and soon thereafter in a monolithic microresonator~\cite{herr_temporal_2014}, they are persistent pulses of light whose remarkable characteristics have attracted attention across the divide of fundamental and applied photonics. On the one hand, temporal CSs have revealed themselves as ideal entities for the systematic investigation of fundamental dissipative soliton physics, permitting controlled experimental insights into a range of nonlinear dynamical phenomena~\cite{leo_dynamics_2013, anderson_observations_2016, anderson_coexistence_2017, lucas_breathing_2017, yi_imaging_2018, jang_synchronization_2018}. On the other hand, they have also enabled -- particularly through their key role in the generation of coherent microresonator Kerr frequency combs~\cite{pasquazi_micro-combs:_2018, kippenberg_dissipative_2018, gaeta_photonic-chip-based_2019} -- ground breaking advances across numerous applications, including all-optical information processing~\cite{jang_temporal_2015, jang_all-optical_2016}, telecommunications~\cite{marin-palomo_microresonator-based_2017,fulop_high-order_2018}, optical frequency synthesis~\cite{spencer_optical-frequency_2018}, detection of extra-solar planets~\cite{suh_searching_2019, obrzud_microphotonic_2019}, spectroscopy~\cite{suh_microresonator_2016, dutt_-chip_2018} and ultrafast optical ranging~\cite{suh_soliton_2018,trocha_ultrafast_2018,riemensberger_massively_2020}.% Here we theoretically and experimentally unveil a new fundamental instability mechanism that allows for the simultaneous co-existence of two distinct \emph{vectorial} CSs.

Temporal CSs have hitherto been predominantly studied in the context of single-component (scalar) systems involving a single (spatial and polarization) transverse mode family of the resonator. It is only very recently that researchers have begun to explore the novel realm of multi-mode (vectorial) systems~\cite{averlant_coexistence_2017, lucas_spatial_2018, bao_orthogonally_2019, nielsen_coexistence_2019, saha_polarization_2020}. In particular, \emph{asymmetric} excitation of two distinct mode families has been shown to allow for the simultaneous support of two non-identical CSs~\cite{lucas_spatial_2018,nielsen_coexistence_2019}, enabling a novel route for the generation of multiple frequency combs from a single device~\cite{lucas_spatial_2018}. However, solitons supported under strongly asymmetric conditions are still effectively \emph{scalar}, being (almost) entirely associated with one of the modes excited. Here, we report on the first experimental study of CSs in a two-mode system under conditions of \emph{symmetric} excitation, and remarkably show that non-identical, \emph{vectorial} CSs can arise via the ubiquitous phenomenon of spontaneous symmetry breaking (SSB). Moreover, we show that appropriate perturbations applied on the cavity driving field allows for one symmetry-broken CS state to be transformed into the other, thus enabling all-optical control and switching of vectorial bit entities.

Our experiments are performed in a macroscopic fibre ring resonator, where the two orthogonal polarization modes of the system are judiciously engineered to be degenerate~\cite{garbin_asymmetric_2020}. We theoretically and experimentally show that, even when the two modes are equally driven, SSB gives rise to two distinct but co-existing CS states with mirror-like, asymmetric polarization states. The soliton symmetry breaking occurs via the same incoherent, Kerr cross-coupling mechanism that was theoretically proposed several decades ago to give rise to SSB of counterpropagating~\cite{kaplan_directionally_1982, hill_effects_2020} or cross-polarized~\cite{haelterman_polarization_1994} \emph{homogeneous continuous wave} (cw) intracavity fields, and that has recently been observed in experiments~\cite{del_bino_symmetry_2017, woodley_universal_2018, garbin_asymmetric_2020, woodley_self-switching_2020}. Our results demonstrate for the first time that such SSB dynamics can also occur for ultrashort, \emph{temporally localised} CS states. We note that the possibility of both temporal and polarization symmetry-breaking in resonators that are synchronously driven with short pulses has also been numerically identified~\cite{copie_interplay_2019}, yet no experimental demonstrations or direct links to CS physics have been presented.

To the best of our knowledge, our results comprise the first experimental observations of spontaneous symmetry breaking of temporal CSs (and non-homogeneous states in general) in a two-component Kerr resonator.  Moreover, whilst SSB has been previously identified and observed for vectorial solitons of conservative systems~\cite{kockaert_stability_1999, cambournac_symmetry-breaking_2002}, and extensively studied theoretically in the context of dissipative systems~\cite{skarka_formation_2014,descalzi_breaking_2020}, the results presented in our work represent the first direct experimental observation of SSB of dissipative solitons in any two-component physical system. As such, our work provides fundamental insights at the intersection of two widely investigated nonlinear phenomena, linking together the rich physics of (vectorial) dissipative solitons~\cite{cundiff_observation_1999, marconi_vectorial_2015, averlant_vector_2016, delque_induced_2007} and spontaneous symmetry breaking. From an applied perspective, our results provide a  novel route for the all-optical storage of topological information, and could have impact on future applications in advanced optical frequency comb generation or sensing using multi-mode resonators.

\section{Results}

\subsection{Theory of CS symmetry breaking}
\begin{figure*}[!t]
 \centering
  \includegraphics[width = \textwidth, clip=true]{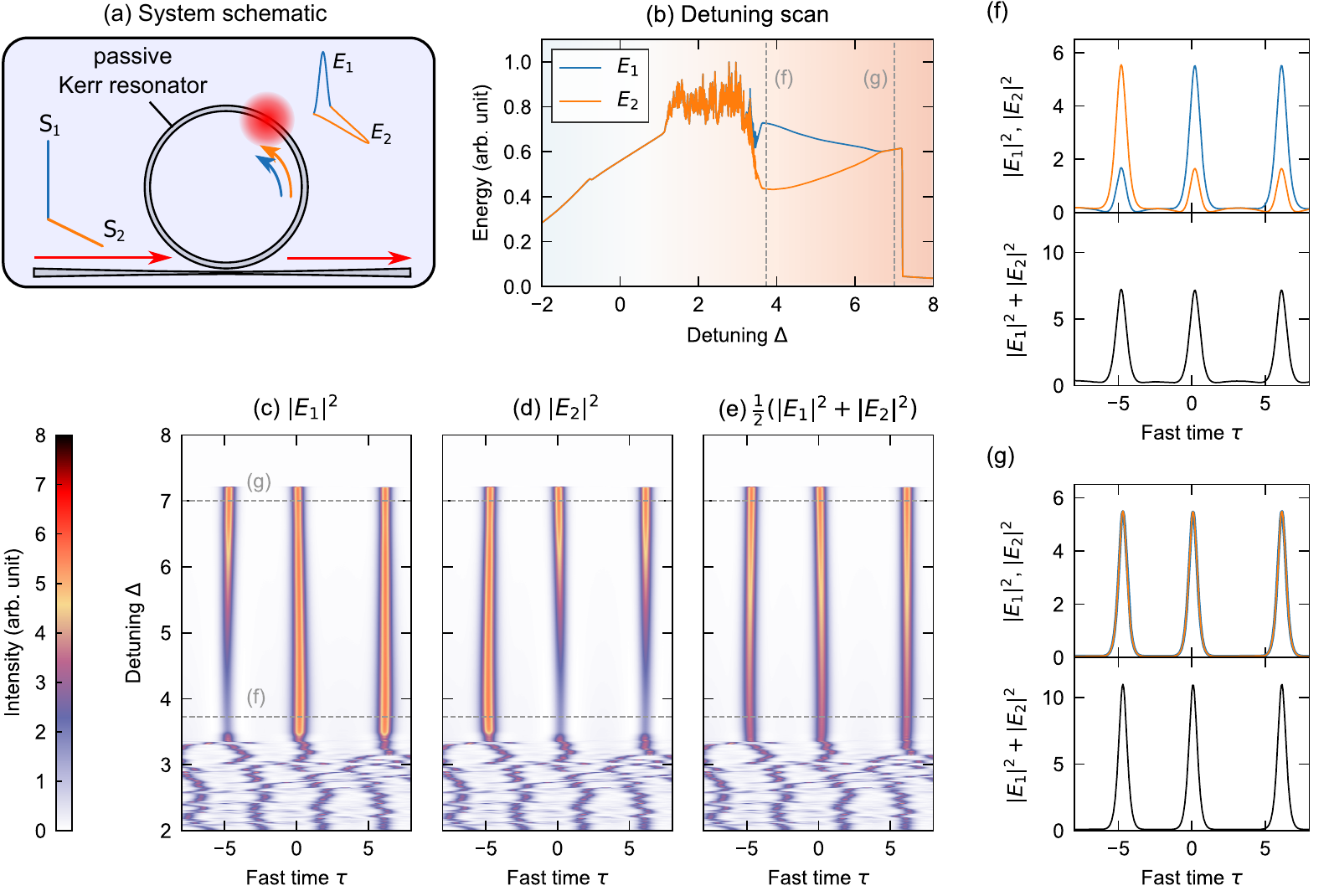}
 \caption{Concept and illustrative numerical simulations. (a) Schematic illustration, showing a passive Kerr resonator whose two orthogonal (polarization) eigenmodes are symmetrically excited. (b) Numerical simulation results, showing the evolution of the integrated intracavity energy contained in cavity mode $E_1$ (blue curve) and $E_2$ (orange curve) as the detuning $\Delta$ is linearly increased. The energies part at the ``soliton step'', revealing SSB. (c, d) Evolution of the intracavity temporal intensity profiles corresponding to modes $E_1$ and $E_2$, respectively, as the detuning is changed; (e) shows the corresponding evolution of the total intensity. (f, g) Top panels show snapshots of the modal intensity profiles at detunings indicated with gray dashed lines in (b)--(e), whilst the bottom panels show the corresponding total intensity. All the results shown were extracted from the same numerical simulation with total driving intensity $X=4.5$ and cross-coupling coefficient $B = 1.6$ similar to the experiments that follow. The colorbar represents the full modal intensities in (c) and (d) and \emph{half} of the total intensity in (e).}
 \label{fig1}
\end{figure*}

\noindent We consider a passive, coherently-driven ring resonator with self-focussing Kerr nonlinearity (anomalous group velocity dispersion) that exhibits two incoherently coupled eigenmodes [see Fig.~\ref{fig1}(a)]. In the mean-field (good cavity) limit, the slowly-varying envelopes $E_{1,2}(t,\tau)$ of the two eigenmodes obey coupled Lugiato-Lefever equations~\cite{geddes_polarisation_1994,averlant_coexistence_2017, nielsen_coexistence_2019, saha_polarization_2020, haelterman_polarization_1994, copie_interplay_2019}. In dimensionless form, the equations read:
\begin{align}
\frac{\partial E_{1,2}}{\partial t} = &\bigg[-1 + i(|E_{1,2}|^2 + B|E_{2,1}|^2 - \Delta_{1,2}) \nonumber \\
&+ i\frac{\partial^2}{\partial\tau^2}\bigg]E_{1,2} + S_{1,2}
\label{LLEs}
\end{align}
Here $t$ is a slow time that describes the field envelopes' evolution at the scale of the cavity photon lifetime, whilst $\tau$ is a corresponding fast time that describes the envelopes' temporal profile. The coupling coefficient $B$ describes the strength of the Kerr cross-phase interaction, $\Delta_{1,2}$ describe the detunings of the driving fields from the respective cavity resonances, and $S_{1,2}$ describe the amplitude components of the driving field along the two eigenmodes. Note that, in contrast to ref.~\cite{copie_interplay_2019}, we consider cw driving such that $S_{1,2}$ are constant scalars. In what follows, we refer to the total driving intensity $X = |S_1|^2 + |S_2|^2$.

We consider the situation where $\Delta_1 = \Delta_2 = \Delta$ and $S_1 = S_2 = \sqrt{X/2}$, such that Eqs.~\eqref{LLEs} exhibit a perfect symmetry with respect to the interchange of the two modes, $E_1\rightleftharpoons E_2$. To probe the intracavity states accessible in experiments, we perform a numerical simulation where the cavity detuning $\Delta$ is linearly increased from small to large values as a function of slow time $t$ (see caption of Fig.~\ref{fig1} for parameters). This simulation mimics the experimental procedure typically used to excite soliton frequency combs in Kerr microresonators by scanning the driving laser frequency across a single cavity resonance~\cite{herr_temporal_2014}. To be closer to experimental conditions, we add uncorrelated white noise on the driving terms $S_1$ and $S_2$ at each integration step. Figure~\ref{fig1}(b) shows the total integrated energy of each intracavity polarization mode as a function of the detuning. The polarization modes initially carry identical intensities, $|E_1|^2 = |E_2|^2$, but as the detuning increases beyond $\Delta>3.2$, the energies part, revealing the breaking of the interchange symmetry $E_1\rightleftharpoons E_2$. With further increase of the detuning ($\Delta > 6.5$), the symmetry is restored.

The symmetry breaking observed in Fig.~\ref{fig1}(b) occurs when the detuning lies within the so-called ``soliton step'' -- a signature widely used to identify the presence of CSs in scalar systems (microresonators in particular)~\cite{herr_temporal_2014}. To gain more insights, Figs.~\ref{fig1}(c) and (d) show the spatiotemporal evolution of the intracavity intensity along the two different cavity modes as the detuning is increased, whilst Fig.~\ref{fig1}(e) shows the corresponding evolution of (half the) total intensity [Figs.~\ref{fig1}(f) and (g) show snapshots at selected detunings as indicated]. As in scalar systems, the intracavity fields initially correspond to homogeneous cw states, but undergo a Turing-like modulation instability that results in the formation of dissipative patterns that fill the entire system. Despite their complex dynamics, we find that the patterned states are (predominantly) identical across the two modes, $|E_1|^2 = |E_2|^2$.

When the detuning increases beyond $\Delta>3.2$, localised CSs emerge from the patterned state. Remarkably, whilst the solitons that emerge carry identical total intensity [Fig.~\ref{fig1}(e)], they come in two distinct flavours with imbalanced ($|E_1|^2\neq|E_2|^2$) modal intensities, corresponding to mirror-like polarization states [Figs.~\ref{fig1}(c), (d), and (f)]. (Note that the splitting of the integrated energies witnessed in Fig.~\ref{fig1}(b) arises due to the different number of solitons spontaneously excited along each mode.) As the detuning increases further beyond $\Delta > 6.5$, the symmetry of the solitons is recovered [Fig.~\ref{fig1}(g)].

\begin{figure}[!t]
 \centering
  \includegraphics[width = \columnwidth, clip=true]{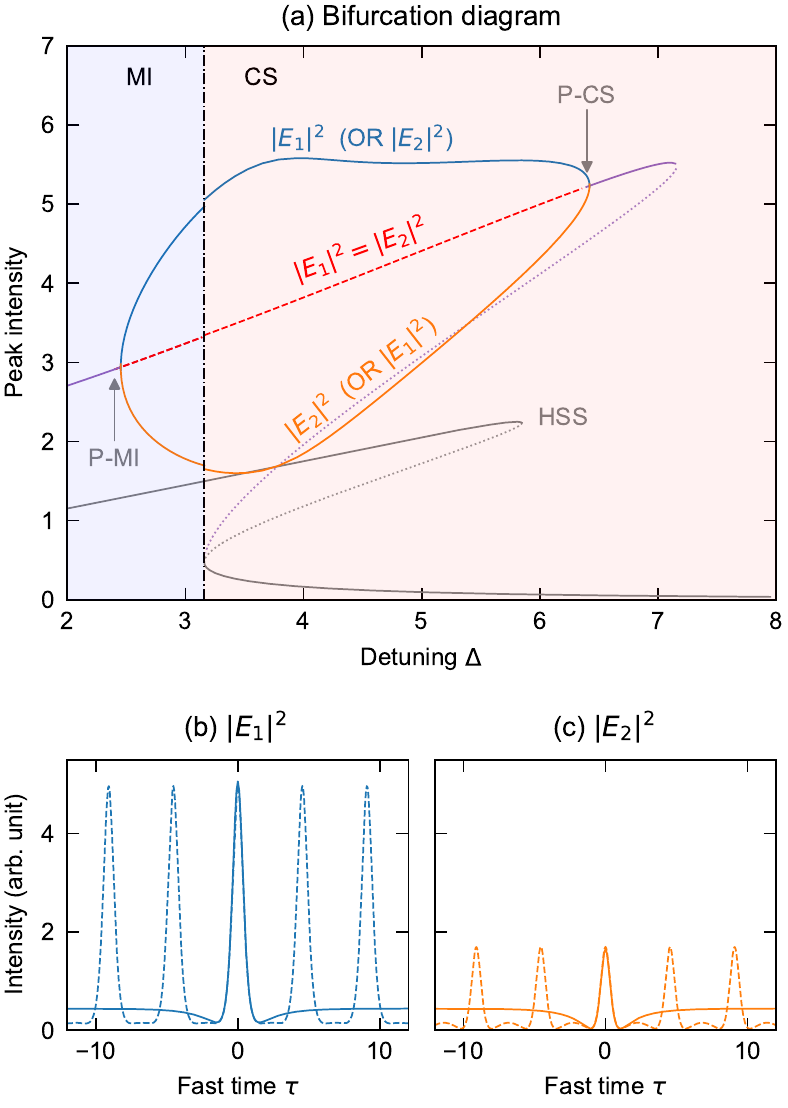}
 \caption{Bifurcations of symmetry-broken CSs. (a) Bifurcation diagram showing the intensity of the homogeneous steady-state (HSS) solutions and the \emph{peak} intensity of the CS and selected modulation instability (MI) pattern solutions of Eqs.~\eqref{LLEs} as indicated [parameters as in Fig.~\ref{fig1}]. The HSS solutions are symmetric at all detunings, whilst the CS and patterned solutions undergo a symmetry breaking pitchfork bifurcation at detunings 6.4 (P-CS) and 2.4 (P-MI), respectively. Solutions that are unstable against polarization symmetry breaking are indicated with red dashed lines. The unconditionally unstable CS and HSS branches are indicated by dotted lines. (b) and (c) compare the modal intensity profiles for a symmetry-broken CS (solid curves) and a patterned solution (dashed curves) that both exist just beyond the up-switching point at $\Delta = 3.16$. }
 \label{fig2}
\end{figure}

To better understand the emergence of asymmetric CS states, we computed the steady-state solutions of Eqs.~\eqref{LLEs} using a Newton-Raphson relaxation algorithm~\cite{mcsloy_computationally_2002}. As shown in Fig.~\ref{fig2}(a), whilst symmetric CS solutions exist over a wide range of detunings, they are stable only for comparatively large detunings. When the detuning decreases, the symmetric soliton loses its stability via a pitchfork bifurcation [P-CS in Fig.~\ref{fig2}(a)], concomitant with the emergence of two stable states with asymmetric, mirror-like modal intensities.  As is the case for the standard scalar CSs of single-mode systems~\cite{coen_universal_2013}, the symmetry-broken CS states exists down to the up-switching point below which the homogeneous state no longer exhibits bistability (here $\Delta  = 3.15$). At that point, the asymmetric soliton branches connect smoothly with branches describing correspondingly asymmetric patterned states. The patterned states originate from the homogeneous state via modulation (or Turing) instability at low detunings (not shown), and whilst they are initially symmetric, their intense peak intensity causes them to undergo their own pitchfork bifurcation [P-MI in Fig.~\ref{fig2}(a)] as the detuning is increased.

\begin{figure*}[!t]
 \centering
  \includegraphics[width = \textwidth, clip=true]{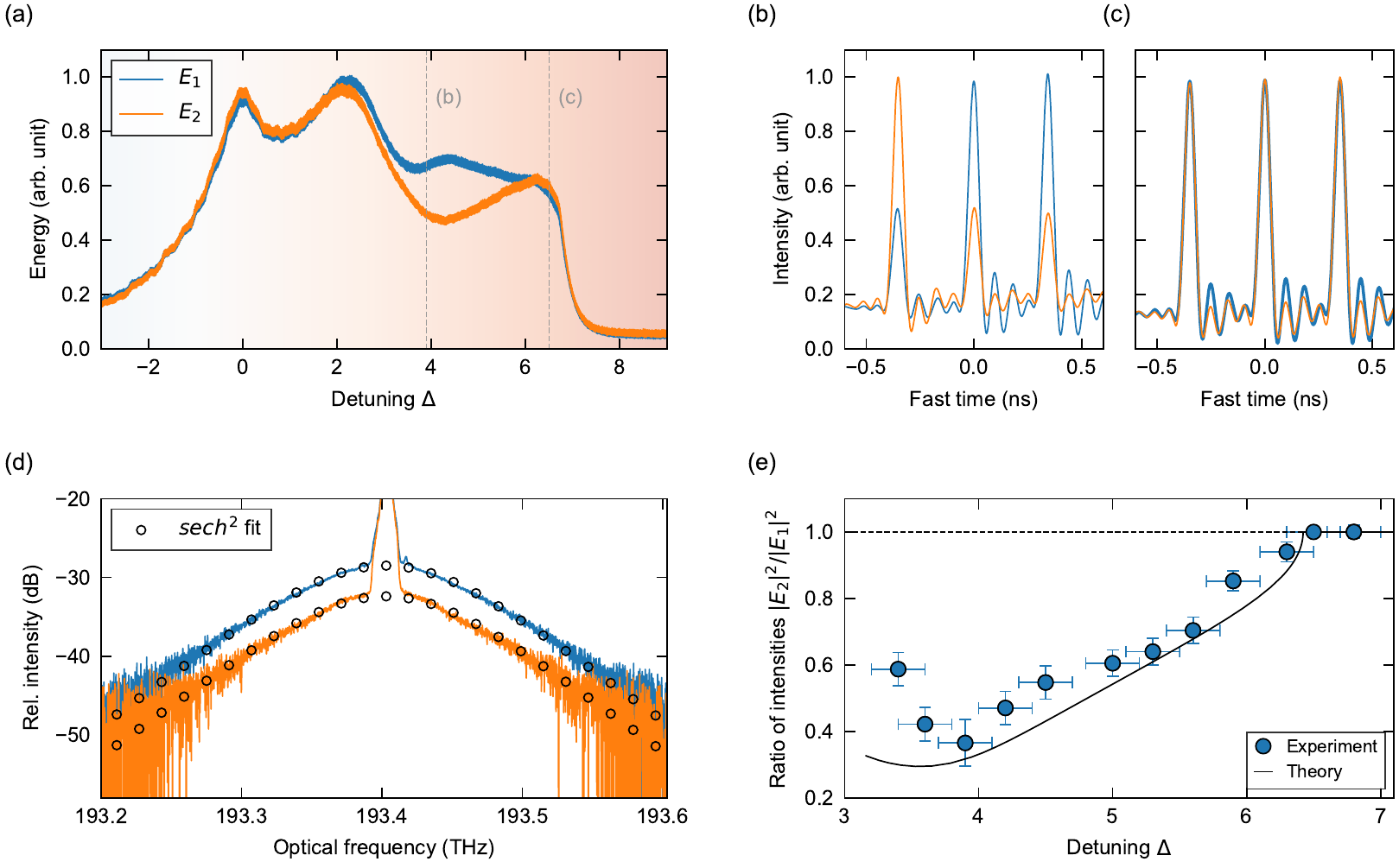}
 \caption{Experimental observation of CS symmetry breaking. (a) Slow photodetector signals measured along the two cavity modes when scanning the laser frequency across a resonance. The experimental parameters are similar to the ones used to obtain the simulation results in Fig.~\ref{fig1}: $X = 4.5$ and $B = 1.6$. (b) Intracavity intensity profiles measured with fast photodetectors (and subsequently sinc-interpolated), showing (b) asymmetric solitons at a detuning of $\Delta = 3.9$ and (c) symmetric solitons at a detuning of $\Delta = 6.5$. (d) Optical spectrum of one of the symmetry-broken CS states. The soliton is predominantly polarised along the ``blue mode'' ($E_1$), but exhibits a small component along the orthogonal ``orange mode'' ($E_2$). Also shown as open circles is the spectrum expected for a 2.7~ps hyperbolic secant CS. (e) Experimentally measured (blue solid circles) and theoretically predicted (black solid curve) ratio of the soliton intensity along the two polarization modes for a ``blue mode'' dominates CS as a function of detuning. The theoretical curve in (e) was obtained from the bifurcation data shown in Fig.~\ref{fig2}(a).}
 \label{fig3}
\end{figure*}

The bifurcation curves shown in Fig.~\ref{fig2}(a) suggest that, in close analogy with the well-known existence mechanism that underpins scalar CSs, the symmetry-broken vector CSs are underlain by the coexistence between a symmetry-broken \emph{patterned} state and a stable \emph{homogeneous background}. We indeed find that the intensity profile of a symmetry-broken CS follows very closely a single period of a symmetry-broken patterned state that exists for the same parameters [see Fig.~\ref{fig2}(b) and (c)]. In this context, we must emphasise that the soliton symmetry breaking does not require a simultaneous breaking of the (corresponding) symmetry of the homogeneous state; results in Fig.~\ref{fig1} and~\ref{fig2} were in fact obtained using a driving power that is below the threshold of SSB of the homogeneous state~\cite{copie_interplay_2019, hill_effects_2020}.

\subsection{Experimental observations}
\noindent For experimental demonstration, we use a setup that is similar to the one used in ref.~\cite{garbin_asymmetric_2020} to observe the symmetry breaking of homogeneous states. There is, however, one key difference: whereas the resonator used in ref.~\cite{garbin_asymmetric_2020} exhibited normal dispersion to avoid CSs or related pattern forming instabilities, our setup uses a resonator made out of standard single-mode optical fibre (SMF-28) with anomalous dispersion $\beta_2 = -20~\mathrm{ps^2/km}$ at the driving wavelength of 1550~nm.

We synchronously drive the resonator with flat-top, quasi-cw pulses with 4.5~ns duration carved from a narrow-linewidth distributed feedback cw fibre laser centred at 1550~nm. To overcome small residual desynchronization between the resonator and the nanosecond driving pulses, we imprint a shallow phase modulation at 2.87~GHz atop the driving field. This phase modulation acts as an attractive potential that traps the solitons~\cite{jang_temporal_2015}, but we have carefully verified that it does not interfere with the SSB phenomenon.

The eigenmodes of interest in our experiments correspond to the two orthogonal polarization modes of the resonator -- they are the principal states of polarization that return to their initial state after one complete round trip. We avoid linear mode coupling between the two modes (which would give rise to phase-sensitive interactions that complicate the cavity dynamics) by driving them with different carrier frequencies spaced by about 80~MHz. As detailed in \cite{garbin_asymmetric_2020}, the carrier frequency shift is chosen to cancel the fibre birefringence such that $\Delta_1=\Delta_2$; additional careful control on the pump levels maintain $S_1=S_2$, allowing us to reliably reach symmetric operating conditions. We project the light output via the 99/1 intracavity coupler along the two polarization modes of the resonator by means of a polarizing beam splitter, and detect each component with both slow (10~kHz) and fast (12.5~GHz bandwidth) photodetectors; the former averages over several cavity round trips whilst the latter gives access to fast temporal dynamics over a single round trip.

Figure~\ref{fig3}(a) shows typical (slow) photodetector traces along each polarization component as the laser frequency is scanned over a single cavity resonance. Here the peak power of the quasi-cw driving field was set to 0.4~W, yielding a normalized driving amplitude $X = 4.5$ that is similar to the simulations above [c.f. Fig.~\ref{fig1}]. In remarkable qualitative agreement with those simulations, we see clearly how the intensities of the two polarization components exhibit a ``bubble'' profile: the intensities are initially identical, part when the detuning approaches the soliton step, and again coalesce as the detuning increases further.

To demonstrate that the system supports two symmetry-broken CSs, we lock the detuning at the soliton step region ($\Delta=3.9$) using the scheme described in ref.~\cite{nielsen_invited_2018}. We then excite solitons by mechanically perturbing the resonator. Figure~\ref{fig3}(b) depicts polarization-resolved oscilloscope traces measured using the fast 12.5~GHz photodetector. The intracavity field is clearly composed of two distinct types of localised structures with asymmetric, mirror-like polarization states. In stark contrast, repeating the measurement at a larger detuning $\Delta=6.5$, we observe the CSs to be completely symmetric [Fig.~\ref{fig3}(c)], in agreement with the theoretical predictions in Figs.~\ref{fig1} and~\ref{fig2}.

The traces shown in Fig.~\ref{fig3}(b) and (c) are limited by the bandwidth of our detection system, and therefore do not reflect the solitons' actual temporal profile. In Fig.~\ref{fig3}(d), we plot the polarization-resolved spectrum of a single symmetry-broken CS (at $\Delta = 3.9$) for which the $E_1$ mode dominates (the $E_2$ component is subdued by about 4.3~dB).  The soliton spectra have sech$^2$ profiles with a 3-dB bandwidth of 0.11~THz along each mode, corresponding to a temporal pulse duration of 2.7~ps. This is in good agreement with the duration of 2.9~ps predicted by numerical simulations of Eq.~\eqref{LLEs} using our experimental parameters. We must emphasise that, while the spectrum shown in Fig.~\ref{fig3}(d) is strictly representative of only one of the two soliton states ($E_1$ dominates), we can also readily observe the other soliton state ($E_2$ dominates) which displays identical characteristics: the only discernable difference between the two solitons is their state of polarization.

\begin{figure*}[!t]
 \centering
  \includegraphics[width = \textwidth, clip=true]{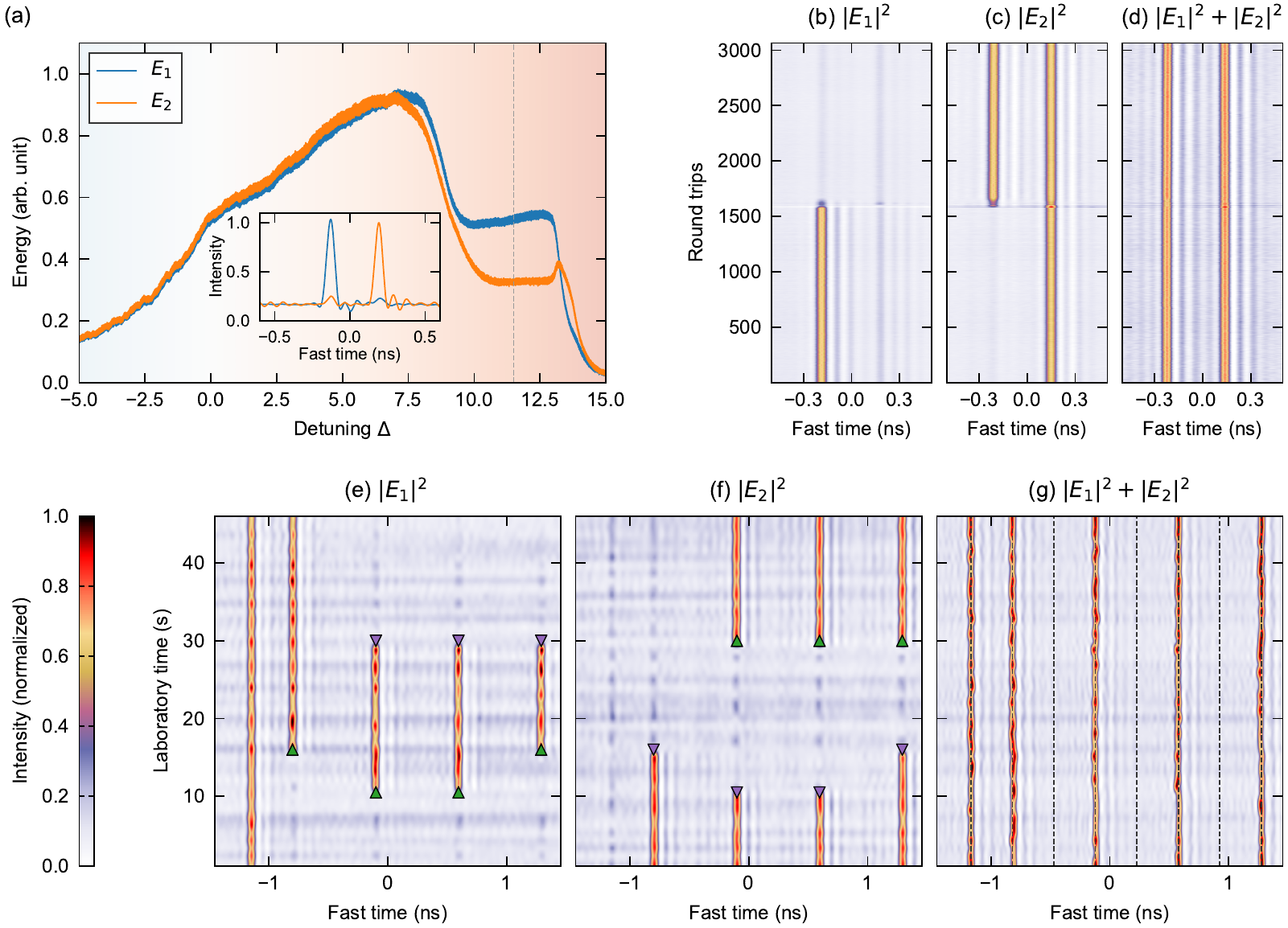}
 \caption{Observation and deterministic switching of symmetry-broken CSs for $X = 21$. (a) Slow photodetector signals measured along the two cavity modes when scanning the laser frequency across a resonance. Inset shows an oscilloscope trace when the detuning is locked at $\Delta= 11.5$ (gray dashed vertical line). (b)--(d) Space-time diagrams, showing how two CSs with different polarization states evolve from roundtrip-to-roundtrip: (b) $E_1$ component, (c) $E_2$ component, (d) total intensity. A polarization perturbation is applied on the driving field at round trip 1,600, and can be seen to enact deterministic switching of a CS from one polarization to the other. (e)--(g) Simultaneous and parallel polarization switching of several CSs:  (e) $E_1$ component, (f) $E_2$ component, (g) total intensity. The green and magenta triangles highlight polarization increase and decrease perturbations, respectively. The colorbar applies to all panels (b)--(g) and the dashed vertical lines in (g) indicate the 2.87~GHz grid defined by the pump phase modulation.}
 \label{fig4}
\end{figure*}

The observations in Figs.~\ref{fig3}(b) and (d) demonstrate that, while the symmetry-broken solitons are predominantly polarized along one of the cavity modes, they also exhibit a noticeable component along the other mode: the solitons are \emph{vectorial}. In good qualitative agreement with the bifurcation curve shown in Fig.~\ref{fig2}(a), our experiments show that the contrast between the two components of the soliton changes with the cavity detuning [blue solid circles in Fig.~\ref{fig3}(e)]. These results were obtained by locking the cavity detuning at different values and by extracting the soliton's modal energies from the detected oscilloscope traces. As can be seen, the solitons are symmetric only at comparatively large detunings, become asymmetric as the detuning is decreased, and remain asymmetric until they cease to exist at $\Delta= 3.15$. Quantitatively, the observed ratios deviate somewhat from the numerical predictions -- especially at low detunings -- which we attribute to imperfect knowledge of experimental parameters and the finite response time of our detection scheme. Nonetheless, these results [together with the results shown in Fig.~\ref{fig3}(b) and (d)] clearly confirm the salient theoretical predictions derived from Figs.~\ref{fig1} and~\ref{fig2}: CSs can undergo SSB in a two-component Kerr resonator, giving rise to coexistence between two distinct, asymmetric vector soliton states. We must emphasize that, being based on a spontaneous symmetry breaking phenomenon, these dynamics are fundamentally different compared to other scenarios where explicitly broken symmetries have been shown to lead to multiplexing of distinct \emph{quasi-scalar} soliton states~\cite{lucas_spatial_2018,nielsen_coexistence_2019,anderson_coexistence_2017, weng_formation_2020}.

The soliton symmetry breaking demonstrated above is not limited to the particular parameters used, but rather occurs whenever the driving power $X$ exceeds a certain threshold (for Kerr cross-coupling coefficient $B = 1.6$ characteristic to our experiment, our simulations show that the threshold $X_\mathrm{t}=3.15$). As an example, Fig.~\ref{fig4}(a) shows photodetector traces measured as the laser frequency is scanned over a cavity resonance with the peak power of the driving pulses set to 1.7~W (corresponding to $X = 21$). The splitting of the modal intensities in the vicinity of the soliton step is evident. We also find that the contrast between the solitons' modal intensities increases with the driving power; for large $X$, the symmetry-broken solitons are aligned almost entirely along one of the cavity modes [see e.g. inset of Fig.~\ref{fig4}(a)], despite both modes being identically driven.

As expected for an SSB instability, the solitons that are excited select their polarization randomly, with the two different polarizations being equally probable. But remarkably, our experiments and simulations show that it is possible to deterministically switch one soliton state to the other via appropriate perturbations. We achieve this in our experiments by using a polarization modulator to locally perturb the cavity driving field. Thanks to the timing reference provided by the phase modulation used to trap the solitons, we are able to selectively address, and hence switch, individual solitons. Our experiments show that a localised reduction in intensity along polarization mode $E_1$ (which corresponds to an increase along mode $E_2$) applied on the driving field at a position with a CS in the $E_1$ mode allows that soliton to be switched onto the other $E_2$ mode (and vice versa).

Figures~\ref{fig4}(b)--(d) show real-time experimental measurements that demonstrate deterministic switching of soliton polarization. Here we plot vertically concatenated sequences of fast oscilloscope traces measured along the two polarizations [Figs.~\ref{fig4}(b) and (c)] as well as the corresponding total intensity [Figs.~\ref{fig4}(d)]. The experiment is initialised with two solitons close to each other: the leading (left) soliton is (predominantly) aligned along mode $E_1$ while the trailing (right) soliton is (predominantly) aligned along mode $E_2$. From the start of round trip 1,600, we apply a polarization perturbation on the cavity driving field that consists of an intensity decrease (increase) along the $S_1$ ($S_2$) component, and that is localised over a 2~ns temporal interval that encompasses both solitons (the perturbation is synchronously applied for about one cavity photon lifetime, or seven round trips). As can be seen, the perturbation causes the leading CS in the $E_1$ mode to switch into the $E_2$ mode. In stark contrast, the perturbation has no effect on the trailing soliton, since an intensity decrease along $E_1$ (increase along $E_2$) does not permit the switch $E_2\rightarrow E_1$. Likewise, we observe no lasting effect at temporal positions where a CS is not initially present. This observation demonstrates that the perturbation is not sufficiently strong to excite new solitons, but rather only allows for the switching of one mirror-like state to the other.

Several solitons can be switched individually and in parallel, allowing for complex all-optical manipulations of polarization-multiplexed information. In Fig.~\ref{fig4}(e)--(g), we demonstrate the manipulation of an 8-bit sequence at 2.87~GHz (defined by our phase modulation) that consists of five solitons (and three empty bit slots). Over the course of the measurement, several of the solitons are switched from one polarization to the other and back. Remarkably, the total intensity remains almost constant during the manipulations [see Fig.~\ref{fig4}(g)], indicating pure polarization dynamics.

\subsection{Discussion}

\noindent We have reported on the theoretical prediction and experimental observation of spontaneous symmetry breaking of temporal CSs in a two-mode Kerr nonlinear ring resonator. The SSB instability gives rise to two coexisting CS states with asymmetric, mirror-like intensity distributions across the two modes of the system. We have obtained clear experimental evidence of such soliton symmetry breaking, and shown that appropriate perturbations enable deterministic switching between the two asymmetric soliton states. Our experimental observations are in very good agreement with numerical simulations.

Whilst beyond the scope of the present Article, our experiments and simulations reveal that the symmetry breaking is not limited to stable CSs, but can also occur for periodically and chaotically oscillating solitons. Our work paves the way for further studies into the rich dynamics of SSB effects in two-component Kerr resonators. From an applied perspective, the possibility to sustain and control two distinct symmetry-broken CS states could enable novel all-optical functionalities, including storage and manipulation of topological information and the realisation of three-valued logic gates. Finally, to the best of our knowledge, our results constitute the first direct experimental observations of spontaneous symmetry breaking of vectorial dissipative solitons in any two-component physical system. As such, our work provides fascinating new connections and insights at the interface of vector solitons, dissipative solitons, and spontaneous symmetry breaking.

\section*{Acknowledgements}
\noindent We acknowledge financial support from the Marsden Fund, the Rutherford Discovery Fellowships, and the James Cook Fellowships of the Royal Society of New Zealand.

\end{document}